\documentclass[twocolumn,showpacs,preprintnumbers,amsmath,amssymb,aps,prl]{revtex4}
\usepackage{graphicx}% Include figure files
\usepackage{dcolumn}% Align table columns on decimal point
\usepackage{bm}% bold math

\begin{document}

\preprint{M. W. Kim \textit{et al}.}

\title{Roles of orbital as a nexus between optical and magnetic properties in cubic $R$MnO$_{3}$ ($R$ = La, Pr, Nd, Gd, Tb)}

\author{M. W. Kim$^{1}$}
\author{S. J. Moon$^{1}$}
\author{J. H. Jung $^{2}$}
\author{Jaejun Yu$^{3}$}
\author{Sachin Parashar$^{1}$}
\author{P. Murugavel$^{1}$}
\author{T. W. Noh$^{1}$}
\email[corresponding author: ]{twnoh@phya.snu.ac.kr}
\affiliation{$^{1}$ReCOE \& School of Physics, Seoul National
University, Seoul 151-747, Korea\\
$^{2}$ Department of Physics, Inha University, Incheon 402-751,
Korea
\\$^{3}$CSCMR \& School of Physics, Seoul National University,
Seoul 151-747, Korea}

\begin{abstract}
We investigated the $ab$-plane absorption spectra of $R$MnO$%
_{3}$ ($R$ = La, Pr, Nd, Gd, and Tb) thin films. As the $R$-ion
size decreases, we observed a drastic suppression of the 2 eV
peak, \textit{i.e.} the inter-site optical transition between
spin- and orbital-aligned states across the Mott gap. We found
that both lattice distortion and the corresponding orbital mixing
of the ordered orbital state should play an important role in the
2 eV peak suppression. We also found that the 2 eV spectral weight
is proportional to the $A$-type antiferromagnetic ordering
temperature, which suggests that the magnetic interaction might be
sensitively coupled to the orbital mixing.
\end{abstract}

\pacs{75.70.-i,77.90.+k,78.20.-e}

\maketitle
LaMnO$_{3}$ has been known as a mother compound of the
colossal magnetoresistance(CMR) manganites, where charge, spin,
lattice, and orbital degrees of freedom interplay with each other
to determine their intriguing physical properties
\cite{Tokura-Science,Tobe1}. LaMnO$_{3}$ has an
orthorhombic structure with four 3$d$-electrons: three $t_{2g}$ and one $%
e_{g}$ electrons. Since three $t_{2g}$ electrons form an orbitally
closed shell, many physical properties are believed to be
determined by its $e_{g}$ electron. In its ground state,
LaMnO$_{3}$ is a Mott-insulator \cite {Kim-NJP,Kovaleva,Inami}
with the $A$-type spin and the $C$-type orbital orderings, which
are schematically drawn in Figs. 1(a) and 1(b), respectively. The
antiferromagnetic (AFM) ordering temperature $T_{N}$ is about 140
K, and the orbital ordering temperature is around 800 K. The
occurrence of the spin- and orbital-ordered state has been
understood in terms of the cooperative Jahn-Teller (JT) transition
\cite{Goodenough}.

Rare-earth substitutions of the La ion provide an intriguing phase diagram
for $R$MnO$_{3}$ ($R$ = rare-earth ion) \cite{Kimura-PRB}, as illustrated in
Fig. 1(c). As the $R$-ion size $r_{R}$ decreases, the crystal structure of $%
R $MnO$_{3}$ changes from orthorhombic ($R$ = La--Dy) to hexagonal ($R$ =
Ho--Lu). TbMnO$_{3}$ and DyMnO$_{3}$ are located near the structural phase
boundary, and they have attracted lots of attention recently due to their
complicated low temperature magnetic states and multiferroic properties \cite
{Kimura-Nature}. On the other hand, the magnetic properties of the
orthorhombic perovskite $R$MnO$_{3}$ ($R$ = La--Tb) has a rather simple $R$%
-dependence: $T_{N}$ decreases with decreasing $r_{R}$. Structural
deformations, such as buckling and distortion of the MnO$_{6}$ octahedra,
also increase. According to the Goodenough-Kanamori rule \cite
{Goodenough,Kanamori}, the orbital overlap of electrons should be crucial in
determining the magnetic interaction. The rule takes into account the
overlap in terms of the Mn-O-Mn bond angle $\phi $. However, the rapid
decrease of $T_{N}$ with the $R$-ion substitution is rather unexpected,
since the change of $\phi $ is less than 10$^{\circ }$. In addition to the
lattice distortion due to the $\phi $ variation, a neutron scattering
measurement showed that the $e_{g}$ electron state should have a mixed
character of $|3z^{2}-r^{2}\rangle $ and $|x^{2}-y^{2}\rangle $ orbitals and
that the degree of orbital mixing varies systematically with $r_{R}$ \cite
{Alonso}. This orbital mixing could also affect the orbital overlap of
electrons. Therefore, $R$MnO$_{3}$ ($R$ = La--Tb) is an ideal system to
investigate roles of the lattice distortion and the orbital mixing in
numerous physical properties.

Optical spectroscopy has been known to be a powerful tool to investigate the
orbital degrees of freedom \cite{Kim-NJP,Kovaleva,Kim-PRL1,Lee}. In this
Letter, we report the $ab$-plane optical responses of epitaxial $R$MnO$_{3}$
($R$ = La, Pr, Nd, Gd, and Tb) films. We find that the spectral weight of
the optical transition across the Mott gap, located around 2 eV, decreases
rapidly as $r_{R}$ decreases. This dramatic reduction of the spectral weight
cannot be explained in terms of the conventional model based on the
structural variations. We demonstrate that the spectral weight change could
be explained by taking account of the structural distortion, \textit{i.e.},
the change of $\phi $, and the orbital mixing. We also find that the
measured spectral weight change is proportional to the variation of $T_{N}$.

High quality $R$MnO$_{3}$ ($R$=La, Pr, Nd, Gd, and Tb) thin films were grown
on double-side-polished (LaAlO$_{3}$)$_{0.3}$(SrAl$_{0.5}$Ta$_{0.5}$O$_{3}$)$%
_{0.7}$ substrates by using the pulsed laser deposition. From
x-ray diffraction measurements, it was found that all the films
grew epitaxially with their $c$-axis perpendicular to the film
surfaces. Details of the film growth and their characterization
were reported elsewhere \cite{Murugavel}. Transmission spectra of
the films were measured from 0.4 to 4.0 eV by using a grating
spectrophotometer. The absorption coefficients were determined by
taking the logarithm of the transmittance, subtracting that of the
substrate, and dividing by the film thickness. Since the
normal-incident
optical geometry was used, the absorption spectra should come from the $ab$%
-plane responses of the films.
%%%%%%%%%%%%%%%%%%%%%%%%%%%%%%%%%%[figure 1]
\begin{figure}[tbp]
\includegraphics[width=3.3in]{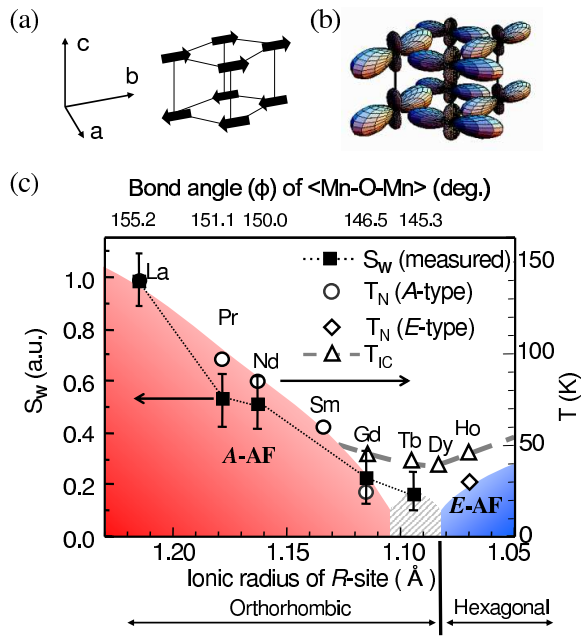}
\caption{(color online). (a) The $A$-type spin and (b) the
$C$-type orbital ordering pattern of electrons at the Mn-sites in
LaMnO$_{3}$. (c) A schematic magnetic phase diagram of
$R$MnO$_{3}$, redrawn from Ref. \protect\cite{Kimura-PRB}. The
spectral weight of 2 eV peak, S$_{W}$ (the solid square), shows
similar $R$-ion size dependence with the $A$-type AFM ordering
temperature ($T$) (the open circles). The $E$-type AFM ordering
$T$ (the open diamonds) and the incommensurate spin ordering $T$
(the open triangles) are also shown. The hatched area represents
the region where the commensurate spin order and ferroelectric
property emerges. } \label{Fig:1}
\end{figure}
%%%%%%%%%%%%%%%%%%%%%%%%%%%%%%%%%%
Figure 2(a) shows the absorption spectra $\alpha (\omega )$ of
$R$MnO$_{3}$ at room temperature, where all the samples should be
in the $C$-type orbital ordered state \cite{Too}. The spectra of
LaMnO$_{3}$ are composed of a peak near 2 eV and much stronger
absorption peaks above 3 eV. The higher energy absorption features
come from the charge transfer transition from O $2p$ to Mn $3d$
\cite{Kim-NJP,Jung}. After long debates, numerous recent
experiments clearly demonstrated that the 2 eV peak should be
interpreted as an inter-site transition across the Mott gap in the
orbitally degenerate Hubbard model (ODHM)
\cite{Kim-NJP,Kovaleva,Inami}. As shown in Figs.\ 1(a) and 1(b),
the correlation-induced transition within the $ab$-plane should
occur between the $e_{g}$ electron states at the neighboring sites
in the ferromagnetic-spin (FM) and antiferro-orbital (AFO)
configuration \cite {Kim-NJP,Kovaleva,Energy}. The absorption
spectra of other $R$MnO$_{3}$ have very similar spectral features.
As $r_{R}$ decreases, $\alpha (\omega )$ for the charge transfer
transition above 3 eV is nearly independent of the $R$-ion,
however, $\alpha (\omega )$ for the correlation-induced 2 eV peak
becomes strongly suppressed. To obtain more quantitative
information, we estimated the spectral weight S$_{W}$ by
subtracting the charge transfer transition background and
integrating $\alpha (\omega )$ from 0.2 to 2.7 eV. The
experimentally determined S$_{W}$, marked as the solid squares in
Fig. 2(b), becomes drastically suppressed with decreasing $r_{R}$.

Such a dramatic decrease of S$_{W}$ is rather unexpected. All the $R$MnO$%
_{3} $ compounds, studied in this work, have the same orthorhombic crystal
structure and the same spin/orbital ordering pattern, but only with a
relatively small variation of $\phi $. Let us look into the possible role of
the structural variations of $R$MnO$_{3}$ in the large S$_{W}$ change.
According to the chemical grip estimate \cite{Harrison}, the inter-site
transition between the $d$ states can vary approximately as $cos^{4}\phi $.
As shown in Fig.\ 2(b), the contribution of the structural variations to the
S$_{W}$ change could be as large as 30$\%$, but is still much smaller than
the experimentally observed S$_{W}$ changes. Therefore, the electronic
structure change due to the structural variation alone cannot explain the
large suppression of S$_{W}$.
%%%%%%%%%%%%%%%%%%%%%%%%%%%%%%%%%%[figure 2]
\begin{figure}[tbp]
\includegraphics[width=3.3in]{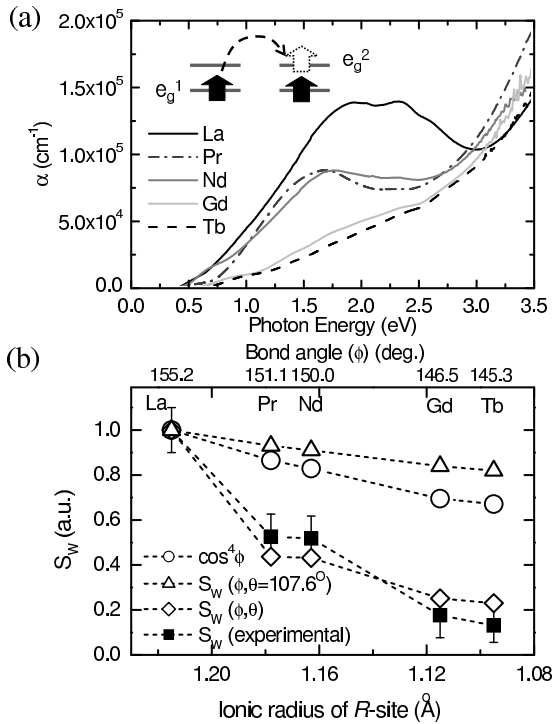}
\caption{(a) Absorption spectra of $R$MnO$_{3}$ ($R$ = La, Pr, Nd,
Gd, and Tb) thin films at room $T$. The inset schematically
represents the inter-site transition corresponding to the 2 eV
peak. (b) The experimental S$_{W}$ (the solid squares) is compared
with the calculation results: the simple estimation of bandwidth
change ($\sim cos^{4}\protect\phi $)(the open circles), our model
calculation for various $\protect\phi $ by using
the orbitals of LaMnO$_{3}$, \textit{i.e.}, at a fixed $\protect\theta $ $%
(=107.6^{\circ })$ (the open triangles), and our model calculation
for various $\protect\phi $ and $\protect\theta $ (the open
diamonds).} \label{Fig:2}
\end{figure}
%%%%%%%%%%%%%%%%%%%%%%%%%%%%%%%%%%
To elucidate the origin of the 2 eV peak spectral change, we applied the
Fermi-Golden rule and evaluate the corresponding matrix element. Within the
electric dipole approximation\cite{Voon}, S$_{W}$ becomes proportional to $%
|\langle \psi _{f}|-i\hbar \nabla |\psi _{i}\rangle |^{2}$. This
matrix element can be approximated by the second order
perturbation, similarly to the superexchange process, through the
oxygen $p$ orbitals:
\begin{equation}
{S_{W}\varpropto |\langle \psi _{f}|\nabla |\psi _{i}\rangle |^{2}\sim |{%
\sum\limits_{\alpha }\langle \psi _{_{f}}|p_{\alpha }\rangle \langle
p_{\alpha }|\psi _{_{i}}\rangle }|^{2}/\Delta ,}
\end{equation}
by assuming that the energy gap $\Delta $ remains almost unchanged\cite{Anderson}. Here $%
|\psi _{i}\rangle $ and $|\psi _{f}\rangle $ represent the
wavefunctions of the initial occupied and the final unoccupied Mn
$e_{g}$ orbitals, respectively, of the transition considered.
$|p_{\alpha }\rangle $ ($\alpha =x,y,z$) represents the oxygen $p$
orbitals which bridge the Mn $e_{g}$ orbitals. For the 2 eV peak,
the matrix element can be estimated by the inter-site transition
from the occupied $e_{g}$ state at one site to the unoccupied
$e_{g}$ state at the neighboring site with the FM/AFO
configuration, as shown in the inset of Fig.\ 2(a).

First, let us consider the contribution of the Mn-O-Mn bond angle
change in the matrix element. As
shown in Fig.\ 3(a), the buckling of the MnO$_{6}$ octahedra in the GdFeO$%
_{3}$ type lattice will cause a decrease in $\phi $. Without the buckling (%
\textit{i.e.} $\phi \simeq $ 180$^{\circ }$), the larger lobe of the $%
|3x_{1}^{2}-r_{1}^{2}\rangle $-type orbital of a Mn$^{3+}$ site is aligned
to face the smaller lobe of the neighboring orbital orthogonally, as shown
in Fig.\ 1(b). As the buckling is turned on, the orbital lobe of an electron
at one site will rotate with respect to that at the neighboring site, which
will result in a reduction in the inter-site hopping amplitude and thereby a
decrease in S$_{W}$. To evaluate the changes in $|\psi _{_{i}}\rangle $ and $%
|\psi _{_{f}}\rangle $ quantitatively, the rotation of the orbitals in the $%
ab$-plane was formulated in terms of the rotational transformation of the
local ($x_{2}$, $y_{2}$) coordinates by $\phi $ with respect to the local ($%
x_{1}$, $y_{1}$) coordinates, as shown in Fig.\ 3(a). Here, the
local $z_{1}$ and $z_{2}$ axis directions are assumed to be the
same. When the orbital wavefunctions of LaMnO$_{3}$ were used, it
was found that the orbital rotation effect on S$_{W}$ is
proportional to $cos^{2}(\pi -\phi )$. As shown in Fig.\ 1(c), the
$R$-ion substitution in $R$MnO$_{3}$ makes $\phi $ vary from
155.2$^{\circ }$ to 145.3$^{\circ }$ \cite{Alonso,Note1}. In Fig.\
2(b), the calculated values of S$_{W}$ are plotted with the open
triangles. It is obvious that the variation in $\phi $ alone
cannot account for the large change in the experimental S$_{W}$.

Now, let us include the orbital mixing contribution. In a cubic MnO$_{6}$
octahedron, two $e_{g}$ orbitals remain doubly degenerate. Under the JT-type
distortion along the $z$-direction, the $e_{g}$ orbitals become split into
two orthogonal orbitals, \textit{i.e.} $|3z^{2}-r^{2}\rangle $ and $%
|x^{2}-y^{2}\rangle $. However, a neutron scattering experiment showed that
the actual occupied $e_{g}$ orbital of $R$MnO$_{3}$ should be a mixed state
of these two orbitals depending on the local distortion of the MnO$_{6}$
octahedron, and further that the degree of the orbital mixing will vary
depending on $r_{R}$ \cite{Alonso}. To include the orbital mixing effects in
Eq. (1), we constructed realistic Mn $e_{g}$ orbitals using the orbital
mixing angle $\theta $: the occupied orbital at site 1 and the unoccupied
orbital at the neighboring site 2 are written as
\begin{eqnarray}
|\psi _{1}^{occ}\rangle &=&\cos \frac{\theta }{2}|3z_{1}^{2}-r_{1}^{2}%
\rangle +\sin \frac{\theta }{2}|x_{1}^{2}-y_{1}^{2}\rangle  \nonumber \\
|\psi _{2}^{unocc}\rangle &=&-\sin \frac{\theta }{2}|3z_{2}^{2}-r_{2}^{2}%
\rangle +\cos \frac{\theta }{2}|x_{2}^{2}-y_{2}^{2}\rangle ,
\end{eqnarray}
where the subscripts in the wavefunctions represent the different
local coordinates. The unoccupied orbital, corresponding to the
final state of the transition, is orthogonal to the occupied
orbital at site 2. To visualize the orbital mixing effects, the
occupied orbitals for three different $\theta $ values are plotted
in Fig.\ 3(b). [Note that the orbital
configuration shown in Fig.\ 1(b) corresponds to $\phi =180^{\circ }$, $%
\theta =108^{\circ }$.] From Eqs.\ (1) and (2), we obtained,
\begin{equation}
{S_{W}\varpropto \{(\sin \theta -\frac{\sqrt{3}}{2})\cos (\pi -\phi )\}^{2}}.
\end{equation}
Using the reported ($\theta $,$\phi $) values from the neutron scattering
experiment \cite{Alonso,Note1}, we can estimate the values of S$_{W}$($%
\theta $,$\phi $) and plot them with the open diamonds in Fig.\
2(b). The estimated S$_{W}$($\theta $,$\phi $) values agree quite
well with the measured S$_{W}$ change, indicating the importance
of the orbital mixing.
%%%%%%%%%%%%%%%%%%%%%%%%%%%%%%%%%%[figure 3]
\begin{figure}[tbp]
\includegraphics[width=3.3in]{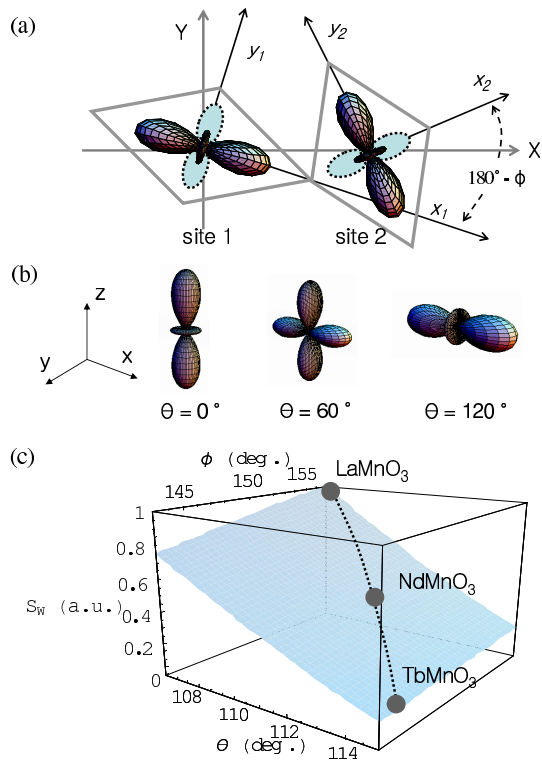}
\caption{(color online). The setup is schematically shown for the
calculation of the orbital mixing angle ($\protect\theta $) and
the Mn-O-Mn bond angle ($\protect\phi $) dependent S$_{W}$. (a)
The local coordinates ($x$,$y$) for the MnO$_{6}$ octahedra and
the Mn $e_{g}$ orbitals (The orbital lobes drawn with the dashed
lines represent the unoccupied orbitals.) (b) The occupied
orbitals of selected $\protect\theta $
values. (c) The calculated S$_{W}$ as a function of $\protect\theta $ and $%
\protect\phi $.} \label{Fig:3}
\end{figure}
%%%%%%%%%%%%%%%%%%%%%%%%%%%%%%%%%%
The $\theta $- and $\phi $-dependence of S$_{W}$ is displayed in Fig.\ 3(c).
Note that $\theta =107.6^{\circ }$ for LaMnO$_{3}$ and $\theta =114.3^{\circ
}$ for TbMnO$_{3}$. Although the variation of the $\theta $ value is about
6.7$^{\circ }$, smaller than that of the $\phi $ value (\textit{i.e.} about
9.9$^{\circ }$), the variation of S$_{W}$ due to the change in the orbital
mixing is larger than that due to the $\phi $ change. A possible reason is
the strong anisotropy in the shape of the orbitals. When $\theta =90^{\circ
} $, the occupied and unoccupied orbitals given in Eq. (3) have the mean
state of the two orthogonal orbitals. As $\theta $ increases, the $%
|x^{2}-y^{2}\rangle $ orbital enhances the wavefunction overlap and the $%
|3z^{2}-r^{2}\rangle $ orbital reduces it within the $ab$-plane, so the
mixing of those two orbitals results in the minimum around $\theta
=120^{\circ }$ in Eq. (3). As shown in Fig.\ 3(c), $R$MnO$_{3}$ are located
near the ($\theta $,$\phi $) space where S$_{W}$ will change rapidly and
depend strongly on $\theta $. Thus, the orbital mixing becomes a crucial
factor in numerous physical properties of $R$MnO$_{3}$, including the change
in S$_{W}$.

The S$_{W}$ values for various manganites are marked with the solid squares
in Fig.\ 1(c). It is remarkable to note that the $R$-dependence of the S$%
_{W} $ change is quite similar to that for $T_{N}$, \textit{i.e.}, the $A$%
-type AFM ordering temperature. With decreasing $R$-ion radius ($r_{R}$)
from La to Tb, $T_{N}$ systematically decreases, \textit{i.e.} from $\sim $
140 K for LaMnO$_{3}$ to nearly zero for TbMnO$_{3}$. Similarly, S$_{W}$
also becomes significantly reduced for TbMnO$_{3}$ as compared with the
value for LaMnO$_{3}$. These similar $R$-dependences of S$_{W}$ and $T_{N}$
are rather surprising, since the S$_{W}$ change comes \textit{from the $ab$%
-plane} response while the AFM ordering at $T_{N}$ occurs \textit{along the $%
c$-axis}.

One possible explanation for this intriguing phenomenon could be an
occurrence of additional FM component in the inter-plane interactions due to
the buckling of the MnO$_{6}$ octahedra. In the undistorted case (\textit{%
i.e.}, $\phi =180^{\circ }$), the $A$-type spin order should occur due to
the FM $e_{g}$-$e_{g}$ interaction within the $ab$-plane and the AFM $t_{2g}$%
-$t_{2g}$ superexchange interaction along the $c$-axis. Here $T_{N}$ is
mainly determined by the latter, since it is much weaker than the $ab$-plane
FM interaction \cite{Hirota}. When the buckling of the MnO$_{6}$ octahedra
occurs, however, the overlap of the $e_{g}$-orbitals between the Mn-planes
brings out a new FM interaction along the $c$-axis. Although $T_{N}$ of $R$%
MnO$_{3}$ should be determined by the competition between the AFM and the FM
interactions, the $r_{R}$ dependence of $T_{N}$ could be realized mostly by
the latter: the AFM interaction should not be so sensitive to the structural
change due to the nature of the orbitally closed $t_{2g}$ levels, but the FM
interaction should critically depend on the buckling. Since the FM
interaction does appear from the tilting of the MnO$_{6}$ octahedra, it
could be closely related to S$_{W}$, which will be proportional to the
square of the electron hopping matrix in the $ab$-plane. This scenario
suggests that the AFM and FM interactions will compete with each other and
achieve a balance around TbMnO$_{3}$. To explain the anomalous magnetic
ground states near the phase boundary, shown in Fig.\ 1(c), Kimura \textit{%
et al.} recently used a two-dimensional anisotropic neighbor
interaction
model \cite{Note2}. Our picture based on a new FM interaction along the $c$%
-axis might provide an alternative starting point to explain the intriguing
magnetic states near the multiferroic phases.

The conventional Goodenough-Kanamori rule takes into account of the orbital
overlap in terms of $\phi $ \cite{Goodenough,Kanamori}. Then, the sign of
the effective magnetic interactions (\textit{i.e.} AFM and FM ground states)
is expected to change near $\phi =135^{\circ }$. In $R$MnO$_{3}$, the GdFeO$%
_{3}$ type distortion can induce the competition between AFM and FM along
the $c$-axis, so one could envisage a disappearance of AFM near $135^{\circ
} $. However, as displayed in Fig.\ 1 (c), the $A$-type AFM order in the
orthorhombic $R$MnO$_{3}$ disappears at $\phi \gtrapprox 145^{\circ }$. This
deviation from the Goodenough-Kanamori rule should originate from the
additional contribution of the orbital mixing to the orbital overlap.

In summary, we reported that the spectral weight of the 2 eV peak changes
drastically with rare-earth ion size in $R$MnO$_{3}$ (R=La, Pr, Nd, Gd, and
Tb). The spectral weight change was successfully understood in terms of the
optical matrix element in which Jahn-Teller distortion and the rotation of
the orbital were taken into account within the orbitally degenerate Hubbard
picture. Similar behaviors between the 2 eV spectral weight and the $A$-type
antiferromagnetic ordering temperature suggest that the superexchange
interaction in $R$MnO$_{3}$ might be tuned by the orbital degree of freedom.

We acknowledge valuable discussions with J. S. Lee, K. W. Kim, S.
S. A. Seo, K. H. Ahn, P. Littlewood, and P. Horsh. This work was
supported by the KOSEF CRI program and CSCMR SRC, and also
supported by the Korean Ministry of Education BK21 project.

%%%%%%%%%%%%%%%%%%%%%%%%%%%%%%%%%%[Refereces]

%%%%%%%%%%%%%%%%%%%%%%%%%%%%%%%%


\begin{references}
\bibitem{Energy} Its energy cost can be estimated as
$U^{\ast}-3J_{H}$, where $U^{\ast }$ is the effective on-site
Coulomb repulsion energy and $J_{H}$ is the Hund's rule coupling
energy. See Ref. \cite{Kim-NJP,Kovaleva}.
\end{references}
\end{document}